\documentclass[useAMS,usenatbib]{mn2e}
\usepackage{graphicx}
\usepackage{latexsym}
\usepackage{amsmath}
\usepackage{subfigure}
\title{Radiation-Hydrodynamic Models of X-Ray \& EUV Photoevaporating Protoplanetary Discs}

\author[J. E. Owen, B. Ercolano, C. J. Clarke \& R. D. Alexander]{J. E. Owen$^1$\thanks{E-mail: jo276@ast.cam.ac.uk}, B. Ercolano$^{1,2}$, C. J. Clarke$^{1}$ \& R. D. Alexander$^3$\\
$^1$Institute of Astronomy, Madingley Road, Cambridge, CB3 0HA, UK\\
$^2$Department of Physics and Astronomy, University College London, Gower Place, London, WC1E 6BT\\
$^3$Sterrewacht Leiden, Universiteit Leiden, Niels Bohrweg 2, 2300 RA Leiden, the Netherlands}
\begin{document}

\pagerange{\pageref{firstpage}--\pageref{lastpage}} \pubyear{2002}

\newcommand{\dd}{\textrm{d}}
\newcommand{\rin}{R_\textrm{\tiny{in}}}
\newcommand{\OO}{\mathcal{O}}
\maketitle

\label{firstpage}

\def\mnras{MNRAS}
\def\apj{ApJ}
\def\aap{A\&A}
\def\apjl{ApJL}
\def\apjs{ApJS}
\def\araa{ARA\&A}

\begin{abstract}
We present the first radiation-hydrodynamic model of a protoplanetary disc irradiated with an X-EUV spectrum. In a model where the total ionizing luminosity is divided equally between X-ray and EUV luminosity, we find a photoevaporation rate of  1.4$\times10^{-8}$ M$_\odot$yr$^{-1}$, which is two orders of magnitude greater than the case of EUV photoevaporation alone. Thus it is clear that the X-rays are the dominant driving mechanism for photoevaporation. This can be understood inasmuch as X-rays are capable of penetrating much larger columns ($\sim 10^{22}$cm$^{-2}$) and can thus effect heating in denser regions and at larger radius than the EUV can. The radial extent of the launching region of the X-ray heated wind is 1-70AU compared with the pure EUV case where the launch region is concentrated around a few AU. When we couple our wind mass-loss rates with models for the disc's viscous evolution, we find that, as in the pure EUV case, there is a photoevaporative switch, such that an inner hole develops at  $\sim$ 1 AU at the point that the accretion rate in the disc drops below the wind mass loss rate. At this point, the remaining disc material is quickly removed in the final 15-20\% of the disc's lifetime. This is consistent with the $10^5$yr transitional timescale estimated from observations of T-Tauri stars. We however note several key difference to previous EUV driven photoevaporation models. The two orders of magnitude higher photoevaporation rate is now consistent with the average accretion rate observed in young stars and will cut the disc off in its prime. 
Moreover, the extended mass-loss profile subjects the disc to a significant period ($\sim$20\% of the disc's lifetime) of  `photoevaporation starved accretion'. We also caution that although our mass-loss rates are high compared to some accretion rates observed in young stars, our model has a rather large X-ray luminosity of $2\times10^{30}$erg s$^{-1}$; further modeling is required in order to investigate the evolutionary implications of the large observed spread of X-ray luminosities in T-Tauri stars.
\end{abstract}


\section{Introduction} 
The structure and evolution of  protoplanetary discs is currently an important and much debated topic in astrophysics. Detailed knowledge of disc evolution is needed to constrain the conditions under which planets form. In particular, an important issue in disc evolution is the mechanism by which the disc is eventually dispersed and the timescales involved, which consequently sets  the time available for planet formation to occur.

\indent For solar type stars it is now well established observationally that at an age of a few Myr a large percentage of stars are still surrounded by discs that are optically thick at infrared wavelengths \citep[e.g.][]{haisch01}. However after $\sim 10^7$yr most stars are observed to no longer be  surrounded by discs.  The relatively small number of discs that are observed to be in transition (between being surrounded by an optically thick disc and having no detectable
disc) implies that the transition occurs on a much shorter $\sim 10^5$yr timescale \citep{kenyon95,duvert00}. This `two timescale' phenomenology  provides a strong constraint on models of disc evolution  and dispersal and is inconsistent with standard viscous-accretion models which predict a power-law decline \citep{hartmann98}\footnote{ It is a matter of current debate whether this two-timescale description also applies to low mass stars \citep[see][]{aguilar08,cornet,currie09}.}.

\indent One of the most successful models in reproducing this `two-timescale' behavior is photoevaporation by radiation from the central star. Recent detection of [Ne~{\sc ii}] emission lines \citep{pascucci07,herczeg07,najita09,flaccomio09,pascucci09} has been interpreted as evidence for irradiation of the disc's atmosphere by energetic photons (extreme ultraviolet or X-ray) and possibly associated photoevaporation \citep{alexander08,pascucci09}. However there is still no consensus over whether photoevaporation is best explained in terms of  far ultraviolet (FUV, $h\nu < 13.6$eV), extreme ultraviolet (EUV, $h\nu > 13.6$eV) or  X-ray irradiation ($h\nu > 100$eV), or some combination of the above. EUV photoevaporation (of optically thick discs) is theoretically well developed \citep{hollenbach94,font04} with a predicted mass-loss rate of $\sim$10$^{-10}$M$_{\odot}$yr$^{-1}$. The mass-loss profile is dominated by  a region within several AU of the characteristic (gravitational) radius $r_g=8.9(M_*/M_\odot)$AU where the internal energy of the gas exceeds the escape potential. Coupling this mass-loss profile to viscous evolution models \citep{clarke01}, this  `EUV-switch' model shows that once the accretion rate through the disc drops to the photoevaporative mass-loss rate, a hole develops and the inner disc drains onto the central star on the  short viscous accretion timescale at a few AU. The inner region is now optically thin to EUV photons and a higher `direct-wind' mass-loss rate of $\sim$10$^{-9}$M$_{\odot}$yr$^{-1}$ is obtained \citep{alexander06a}.  The disc then photoevaporates from the inside out on a timescale of $\sim$10$^{5}$yr \citep{alexander06b}. The relevance of the EUV-switch model, however, depends, albeit weakly,
on the EUV luminosity of the star; the canonical mass loss rates quoted above 
correspond to an  EUV luminosity of 10$^{41}$ photons s$^{-1}$. This cannot be verified by direct observation owing to the strong absorption of ionizing
photons by intervening neutral hydrogen. While \citet{alexander05} derived EUV fluxes from young solar type stars which are consistent with those required  by photoevaporation, it has been pointed out \citep[e.g.][]{glassgold04,ercolano09,glassgold09} that the EUV photons emitted by the star do not necessarily reach the disc, given that EUV photons are extremely easily absorbed, by small columns of neutral material surrounding the young star. X-ray observations of young T-Tauri stars show that neutral hydrogen columns of up to 10$^{22}$cm$^{-2}$ are common around many objects, probably due to accretion funnels, implying that EUV photoevaporation is unlikely to be effective until late times. 

 FUV driven neutral flows have also been considered as a possible mechanism for disc dispersal. In rich cluster environments, \citet{johnstone98} showed  that FUV radiation from nearby stars can dominate the photoevaporation rate. This, however doesn't explain the rapid dispersal of discs around young stars with no nearby source of FUV flux. \cite{gorti09} have considered FUV driven flows from the central star and argue that mass-loss rates of order $10^{-8}$M$_{\odot}$yr$^{-1}$ can be obtained at large radius ($>100$AU). 

\indent In contrast to  EUV radiation, X-rays are able to penetrate larger columns of neutral material. Also X-rays are largely unaffected by accretion funnels (since they are emmitted in the magnetosphere, above the accretion funnels); furthermore the X-ray luminosity function of young T-Tauri star is well known \citep{preibisch,colombo07}. 
For these reasons several authors have considered X-ray irradiation as a possible driving mechanism for photoevaporation \citep[e.g.][]{alexander04,gorti09,ercolano08,ercolano09}. While \citet{alexander04} and \citet{gorti09} quantitatively disagree on the mass-loss rates due to X-ray irradiation, they both conclude that X-ray heating is considerably less important  than EUV heating. \citet{ercolano08,ercolano09} obtained mass-loss rates in the range $10^{-8}-10^{-10}$M$_{\odot}$yr$^{-1}$ and argue that the X-ray heating will dominate the mass-loss and hence the evolution of the disc. While the hydrodynamics of EUV photoevaporation have been solved in detail, there are no prior hydrodynamical studies of X-ray irradiated gas.  \citet{alexander04,gorti09} both use a 1+1D description of the radiative transfer, and \citet{ercolano08,ercolano09} uses a 3D approach; however, in all previous studies the disc structures are assumed to be in {\it hydrostatic equilibrium} and the mass loss-rates are obtained either by assuming a sonic flow from the location where the local gas temperature equals the escape temperature \citep[the $\rho c_s$ method, e.g.][]{alexander04,ercolano08,ercolano09}, or by coupling a spherical Parker wind solution to the hydrostatic structure \citep{gorti09}.  While these methods may (or may not) 
give a reasonable  order of magnitude estimate of the total mass-loss rate,  they are unlikely to give a reliable radial  profile of  the mass-loss rate,  as is needed to determine the evolution of viscously evolving photoevaporating discs.

 In this paper we couple the radiative transfer models of \citet{ercolano09} to two-dimensional hydrodynamical models to obtain a self consistent flow solution for a primordial disc (i.e. an optically thick disc extending to the dust destruction radius) and for inner hole discs (i.e. gas discs with cleared inner holes). In section 2 we present our approach to the radiation-hydrodynamics. In sections 3 and 4 we apply these methods to a primordial disc and discs with inner holes and present the basic results. In section 5 we show the results of various tests we conducted. In section 6 we present the results of
coupling our wind profiles to a viscous evolution model. In section 7 we compare our work to previous results and discuss the model's limitations in section 8. In section 9 we discuss the observational consequences of our model and summarise our main findings in section 10.

\section{Methods}
In what follows we undertake 2D hydrodynamic calculations in which -
under the assumption that the gas is in radiative equilibrium - we
parametrize the gas temperature as a function of local variables. This
parametrization is based on the results of 3D Monte Carlo radiative
transfer calculations using {\sc mocassin} \citep{moc1,moc2,moc3}.
We then verify {\it a posteriori} that the cooling timescale is less than
the flow timescale and hence justify our assumption of radiative
equilibrium. Below we set out the details of the radiative and
hydrodynamic elements of the calculation.

\subsection{Radiative Transfer}\label{mocassin}

\citet{ercolano08,ercolano09} employed the three-dimensional Monte Carlo photoionization and dust radiative transfer code, {\sc mocassin} \citep{moc1,moc2,moc3}, to calculate the temperature and ionization structure of a typical T-Tauri disc irradiated by chromospheric X-ray and EUV flux. {\sc mocassin} uses a stochastic approach to the transfer of radiation and allows for arbitrary geometries and density distributions. The version of the code used for these models was modified by \citet{ercolano08,ercolano09} and we refer to these articles for a detailed description of the code and model setup. We use the model labeled FS0H2Lx1 by \citet{ercolano09} and summarise here only the basic input parameters. 
 \subsubsection{Intial Disc Configuration}
 The initial density structure of a protoplanetary disc surrounding a 0.7~M$_{\odot}$ star (T$_{eff}$ = 4000~K, R$_{*}$ = 2.5~R$_{\odot}$) was taken from the \citet{d'alessio01}  set and was chosen as the model that best fit the median SED in Taurus. This model is in vertical hydrostatic equilibrium in the case that the dust temperatures are set by
irradiation by the optical radiation from the central star and under the assumption of full thermal coupling
between the gas and dust.  The total mass in the disc is 0.026~M$_{\odot}$ with an outer radius of 500~AU. 

 This disc structure,
which provides our starting density distribution, employs a
bimodal dust distribution, where atmospheric dust follows the
standard MRN model but with a very low dust to gas mass ratio (2.5$\times 10^{-5}$ for
graphite and 4.0$\times 10^{-5}$ for astronomical silicates)
and interior dust consists of larger grains with a size distribution
still described by a power law of index -3.5, but with $a_{min}$ = 0.005 $\mu$m and
$a_{max}$ = 1 mm and a dust-to-gas mass ratio of 3.1$\times~10^{-2}$ for graphite and 5.0$\times~10^{-2}$ for
astronomical silicates.
The transition between atmospheric and interior dust occurs at a height
of 0.1 times the midplane gas scale height.

High-energy dust absorption and scattering coefficients are
calculated from the dielectric constants for graphite and silicates
of \citet{laor93}, which extend to the X-ray domain.
Spherical grains are assumed and we use the standard Mie
scattering series expansion for complex refractive indices,
$x|m| < 1000$, where $x = 2 a/\lambda$ is the scattering parameter \citep[see][]{laor93}. For $x|m| > 1000$ and $x|m - 1| < 0.001$
we use Rayleigh-Gans theory \citep{bohren83}, and
for $x|m| > 1000$ and $x|m - 1| > 0.001$ we use the treatment
specified by \citet{laor93}, which is based on geometric
optics.

\subsubsection{Method}

\begin{figure}
\centering
\includegraphics*[width=\columnwidth]{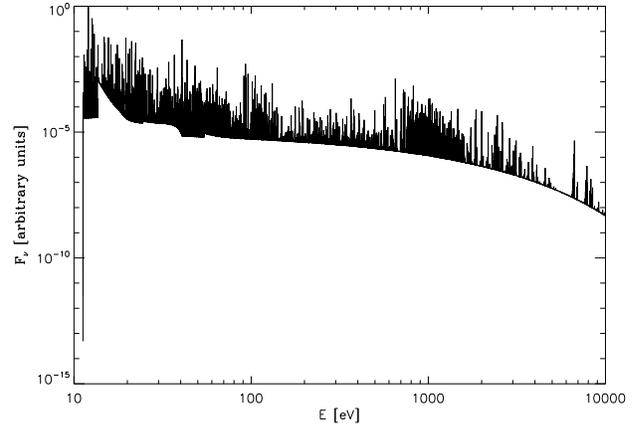}
\caption{Input spectrum used in {\sc mocassin} calculation. See text and \citet{ercolano09} for details.}\label{spectrum}
\end{figure}

This initial
 disc was then irradiated by a synthetic coronal spectrum of X-ray luminosity L$_X$~=~2$\times$10$^{30}$erg s$^{-1}$ (see Figure \ref{spectrum}
) and total ionizing luminosity of L$_{tot}$~=~4$\times$10$^{30}$erg s$^{-1}$. This synthetic
spectrum was a thermal spectrum generated by the plasma code of \citet{kashyap00}; the emission measure
distribution is  based
on that derived  for RS CVn type binaries  by \citet{sanz02} (which peaks at $10^4$K) and fits to Chandra spectra of T Tauri stars by
\citet{maggio07} ( which latter peaks at around log$T=7.5$K). Our use of the input spectrum shown in
Figure \ref{spectrum} implies that we assume a zero column of neutral material close to the source. The
experiments described in \citet{ercolano09} however show that screening columns of up to $10^{20}$
cm$^{-2}$, which would only absorb the EUV component of the spectrum, have little effect on the resulting
photoevaporation estimates; we thus expect our results to generalise to the case of moderate screening
($< 10^{20}$ cm$^{-2}$) since this has little effect on the X-rays in the keV range which dominate the disc
heating (see below). As described in \citet{ercolano09}, the disc temperatures are updated in response to
irradiation by the X-EUV spectrum and the density structure updated so as to attain a situation of
hydrostatic equilibrium for this updated temperature profile. These temperature and density distributions
are then updated iteratively until a situation of both thermal and hydrostatic equilibrium is attained. It
is these converged hydrostatic models 
that form the basis of the parametrization described below, even though the hydrodynamical
calculation evolves away from this initial static structure towards a steady flow solution. We however
check retrospectively that our parametrization also applies to the evolved structure (see Section 5).  

Under the assumption that the thermal and hydrodynamical timescales are decoupled\footnote{The thermal timescale in the
flow is typically less than a year and thus more than an order of magnitude less than the hydrodynamic time scale.}
a coupling of the {\sc mocassin} code with a hydrodynamical code is trivial in principle. It simply consists of feeding the density fields produced by the hydrodynamical code into {\sc mocassin} to calculate gas temperatures that are then in turn fed back to the hydrodynamical code. While conceptually easy, this type of coupling is  computationally prohibitive. 

\begin{figure}
\centering

\subfigure[Relation for Primordial Disc]{\label{fig:relationsa}
\includegraphics*[width=\columnwidth]{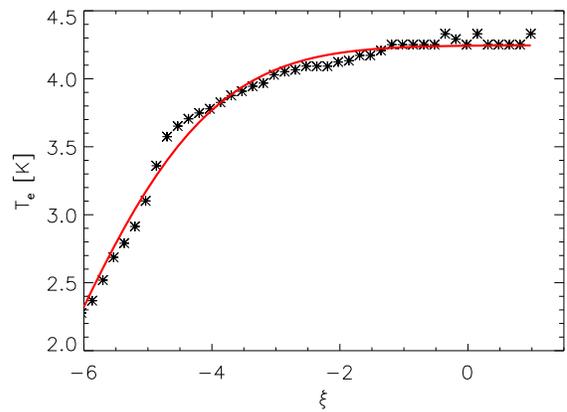}}
\subfigure[Relation for Inner Hole Discs]{\label{fig:relationsb}
\includegraphics*[width=\columnwidth]{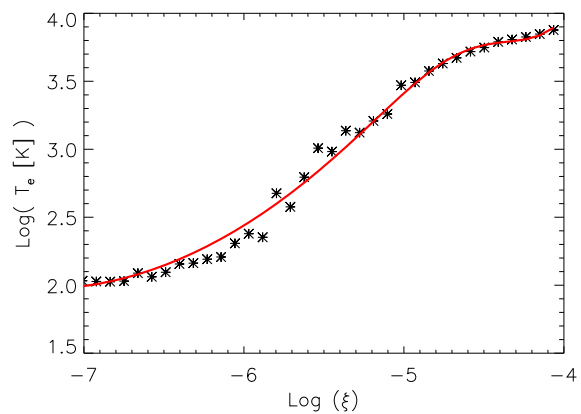}} \caption{Plot of Temperature as a function of ionization parameter for FS0H2Lx1 model \citep{ercolano09} with numerical parametrization shown in red. The black asterisks show the mean temperature in the ionization parameter bins obtained from {\sc mocassin}, with a scatter about the mean of $\sim 0.3$ dex.}\label{fig:relations}
\end{figure}

\indent In order to render the calculation tractable, we exploit  the fact that the wind will be optically thin to high energy photons. In the density and temperature regimes expected in a photoevaporative wind, radiative heating and ionization compete with collision-initiated cooling and recombination and hence the ionization equilibrium strongly depends on the ionization parameter $\xi = L_{\tiny X}/nr^2$ \citep[see][]{tarter69}. The ionization parameter can be conceptually considered as the ionizing energy available to a gas particle at a distance $r$ from the source and will determine the ionization state of the gas and hence the gas temperature at that point \citep[e.g.][]{idea98}.  We show in Figure~\ref{fig:relations} the mean relation between ionization parameter and temperature in the X-ray heated region of
the hydrostatic models of \citet{ercolano09} (upper panel), where we define the X-ray heated region as comprising
gas for which the absorbing  column to the central star is less than $10^{22}$ cm$^{-2}$
(roughly the penetration depth of 1~keV photons). The scatter of the {\sc mocassin} grid cells about the relation
shown is small ($\sim 0.3$ dex)  thus showing  that the ionization parameter is indeed a good determinant of
the gas temperature.  This confirms  the central importance of X-ray heating in determining the thermodynamics
of the heated atmosphere (note that a tight relationship between ionization parameter and temperature is not
expected in general in the case of heating dominated by EUV radiation). In Section 4 we will consider the case
of photoevaporation from discs containing inner holes for which we need to improve our parametrization in the
lower temperature regime.   
The lower panel of Figure~\ref{fig:relations} depicts  
the result of a similar calculation of a disc which is truncated at an inner radius of $16$ AU; we have
confirmed that  this provides also a good fit to models with different inner hole sizes and hence use
this relation in all the hydrodynamic models in Section 4. Note the similarity of the relations for the identical ionization parameter range. This implies that it is the direct field that is dominating in both cases. This should be 
contrasted  with the pure EUV case, where the diffuse field dominates for the optically thick primordial disc and the direct field dominates for discs with inner holes. 

 Since  these relations show that the gas temperature varies monotonically with ionization parameter in the Xray heated regime, we can thus use the ionization
parameter (which depends only on the incident flux and local number density)
in order to prescribe the gas temperature for regions lying at columns
of $< 10^{22}$ cm$^{-2}$ from the central source.

\indent For columns larger than $10^{22}$cm$^{-2}$ \citet{ercolano09} (ECD09) found that the gas and dust were thermally coupled. Hence we use the dust temperatures of \citet{d'alessio01} to fix the gas temperatures for columns greater than $10^{22}$cm$^{-2}$. In order to check that our temperature parametrization is appropriate for the final converged flow solution, we feed back this density field into  {\sc mocassin} and compare the temperatures thus obtained; these results are presented in \S\ref{test}.

\subsection{Radiation-Hydrodynamic Methods}\label{hydro}
In order to accurately determine the mass-loss rates and the kinematic and morphological structure of the evaporative flow, we need to compute the temperature distribution in the flow and feed this into a hydrodynamic code.  We have modified the publicly available {\sc zeus-2d} hydrodynamic code \citep{zeus2da,zeus2db,zeus2dc} to include the heating from X-ray and EUV radiation, by  writing  a new module  that calculates the gas temperature using the methods discussed in \S\ref{mocassin}. We then use this temperature to update the internal energy density of gas ($u$) in the wind via the ideal equation of state:
\begin{equation}
u=\frac{k_B}{\mu m_h(\gamma-1)}\rho T
\end{equation} where $k_B$ is the Boltzmann constant, $m_h$ is the hydrogen mass and $\gamma$ is the ratio of the specific heats. The mean molecular weight is fixed in the flow, to a value of $\mu=1.37125$. The internal energy is only required to update the pressure needed for evaluation of the momentum equation. The pressure is thus determined again from the ideal gas law, via:
\begin{equation}
p=(\gamma-1)u
\end{equation} and hence the procedure is independent of choice of $\gamma$.  This thermal update is done during both the source step and transport step in order to minimise the time to a converged solution.

\indent We assume azimuthal and mid-plane symmetry and since we expect a photoevaporative flow to be approximately spherical at large radius, we use a spherical grid. We neglect  magnetic fields and self- gravity.  The rotation option in {\sc zeus-2d} is enabled, which introduces the necessary pseudo-forces arising from rotating frames. The inner and outer radial boundaries are set to outflow, while the angular boundaries are given the appropriate symmetry boundary conditions. We choose to use the standard van-Leer (second-order) interpolation scheme and the von-Neumann \& Richtmyer parametrization for the artificial viscosity with $q_{\textrm{visc}}=2.0$. 
\section{Primordial Disc}

In order to determine the flow structure for the disc for the majority of its standard viscously accreting phase we run a single hydrodynamic model. We expect the structure of the flow to be set by the temperature of the
X-ray heated region and, to a lesser extent, by the temperature of the disc atmosphere which is controlled
by optical heating of dust in the disc's surface layers. The structure of the underlying disc is insensitive
to this provided that it is very optically thick, both in the X-ray and optical regions. This is amply fulfilled by
primordial (untruncated) discs over many orders of magnitude in accretion rate, thus justifying the
fact that we use a single hydrodynamic model to evaluate the wind mass loss profile for the entire period
of the disc's evolution prior to gap opening (see Sections 4 and 6). A similar argument has previously been
applied also to the EUV case, where the mass loss profile is unaffected by the total column in the disc,
provided it is optically thick in the EUV \citep{hollenbach94}.  

\subsection{Numerical Method}
The grid is chosen to be in the range $\theta=[0,\pi/2]$ and $r=[0.33,100]$AU\footnote{Note throughout this paper we use $(r,\theta,\phi)$ and $(R,\varphi,z)$ to distinguish between spherical and cylindrical polar co-ordinates respectfully.} with 100 grid cells in the angular direction and 250 in the radial. Grid cells are logarithmically spaced in the radial direction, so  that we have adequate resolution at small radius: in particular we ensure that we
  can  resolve the onset of any flow and can resolve the scale height of the X-ray `dark' disc at all radii.
 As initial conditions,  the vertically hydrostatic density and temperature structures generated by \citet{ercolano09} were  interpolated onto the {\sc zeus-2d} grids and assigned Keplerian rotation (suitably adjusted for the small radial pressure gradient).  The disc structure was then allowed to evolve hydrodynamically using the modified code, with the gas temperatures being updated as described in \S\ref{mocassin} at every timestep,  until a `steady-state\footnote{Where the term `steady-state' strictly means quasi-steady-state or more formally $\dot{M}_w \times t \ll M_\textrm{\tiny disc}$.}' disc/wind system was obtained. It should be noted that the density structure of the X-ray `dark' regions 
is  also allowed to evolve hydrodynamically, i.e. it is
not reset to the original density structure at every time-step (compared with  previous EUV hydrodynamic simulations e.g. \citealt{alexander08,font04} which - since they did not model the EUV `dark' disc - reset the
 base density of the flow to  its  original value at each time-step).
We emphasise that the  mass loss over the duration of the simulation is  much less than the initial disc mass.

\indent The evaporative flow was found to have converged after $\sim$4-6 orbital timescales at the outer boundary (convergence was also checked by determining the Jacobi potential was conserved along streamlines: See \S\ref{test}). The flow structure after 10 orbital timescales at the outer boundary is used in our analysis.  

\subsection{Results}
\begin{figure}
\centering
\subfigure[Initial Disc]{\label{fig:flowb}\includegraphics*[width=\columnwidth]{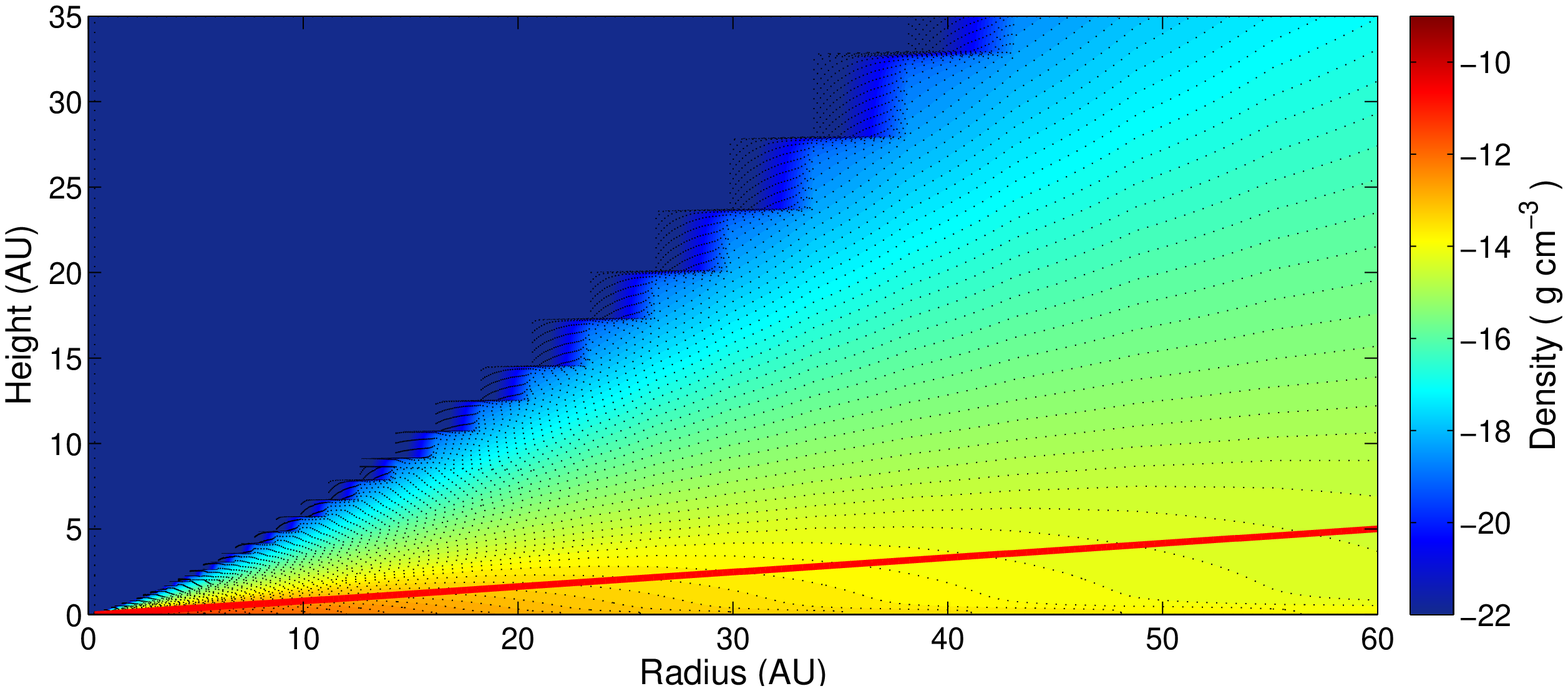}}
\subfigure[Flow Solution]{\label{fig:flowa}\includegraphics*[width=\columnwidth]{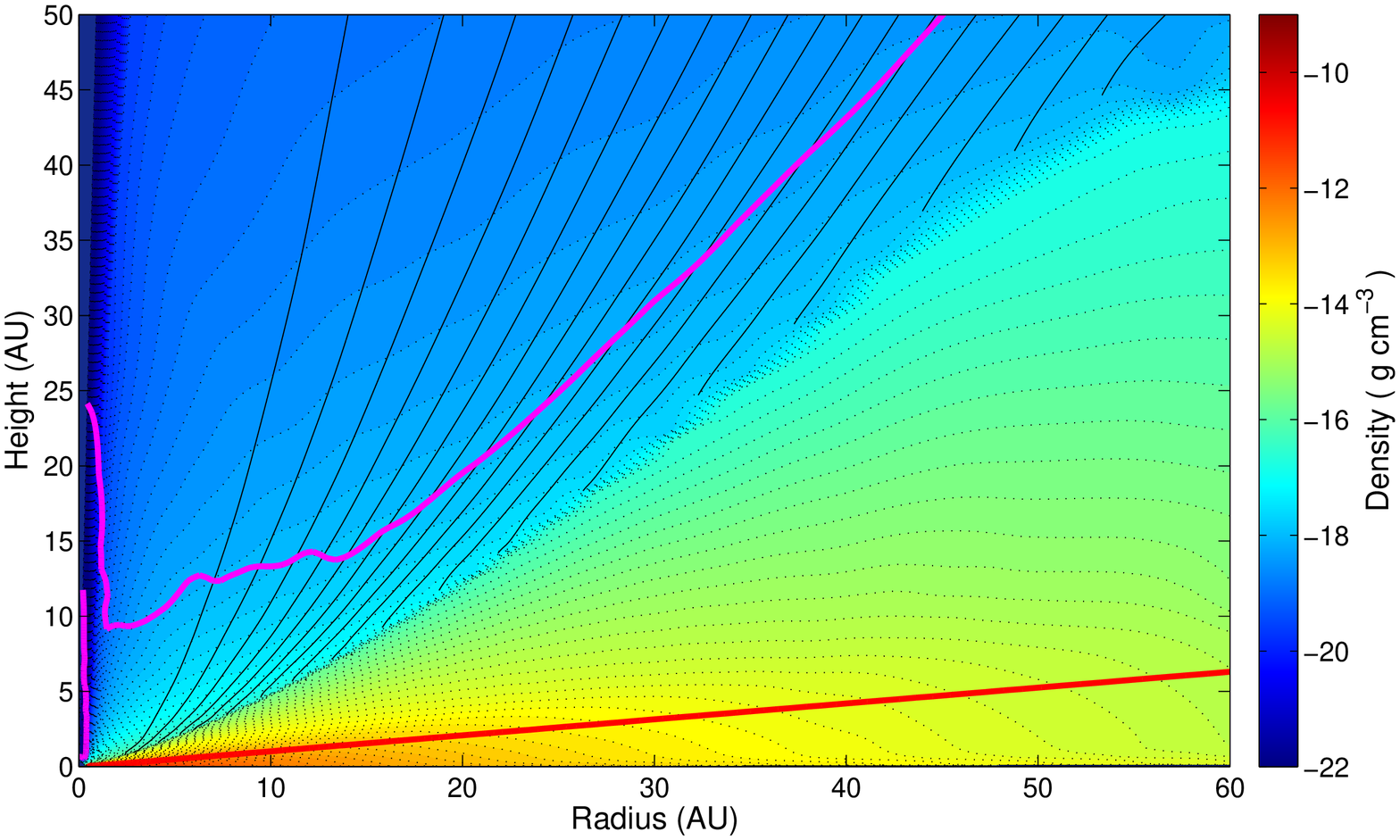}}
\caption{The density structure of a primordial disc; with the initial disc shown in (a) and the 'steady-state' flow solution in (b). The red line indicates the $\tau$=2/3 surface from the central star in both plots. In (b); streamlines are plotted at 5\% intervals of the integrated mass loss rate, with the sonic surface plotted in magenta. Note the smoothness
of the sonic surface indicates that we are extremely close to a true `steady-state'. The snapshot shown is at a simulation time of 10 orbital periods at 100AU.}\label{fig_disc}
\end{figure} 

In Figure \ref{fig_disc} we show the density structure of the wind and underlying primordial disc with streamlines overlain. The Figure shows that the flow originates near the point at which the X-ray flux can penetrate, i.e. a column of $10^{22}$cm$^{-2}$.  This also indicates that the flow launching surface is significatly higher than the disc's photosphere as seen by the star (red solid line in \ref{fig:flowa}, where we have used the opacities of \citealt{d'alessio01} to determine this surface). Also, comparing the intial disc structure (Figure \ref{fig:flowb}) and the final flow solution (Figure \ref{fig:flowa}) it is clear that the optically thick disc region is very similar and is thus observationally consistent with the SED of  classical T-tauri stars by construction, with the surface of optical depth to stellar photons of $\tau=2/3$ lying at $\sim$ 0.1$R$. We stress that although the X-ray heated wind launches from well above the
disc mid-plane, both the wind and the upper regions of the underlying disc
are
optically thin to stellar photons and thus the fraction of young stars
that
are optically obscured by their discs remains at the $\sim 10 \%$ level
implied by the \citet{d'alessio01} models. 

\begin{figure}
\includegraphics[width=\columnwidth]{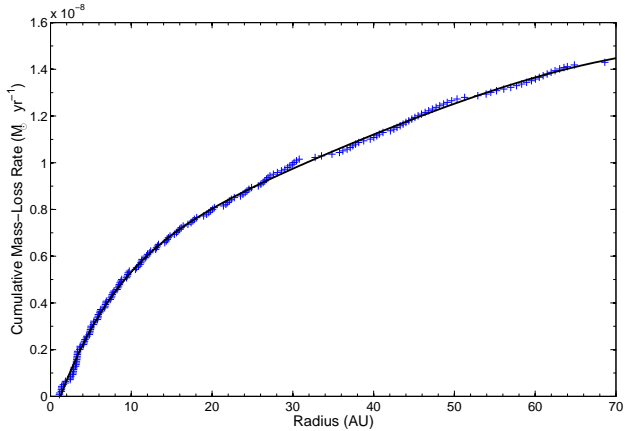}
\caption{Cumulative mass loss rate as a function of cylindrical radius, with the numerical fit used in the viscous evolution shown as the solid line. Note the smooth profile with only one `bump', indicates the flow is extremely close to a true `steady-state'.}\label{fig_massl}
\end{figure}

\indent Figure \ref{fig_disc} also shows that the mass loss is dominated from the region  5-20 AU, although there is still significant mass-loss at larger radii. We note that the streamlines are close to radial at large radius similar to those obtained from pure EUV models \citep{font04,alexander06a,alexander08}. The observation that the flow is approximately radial at large radius allows us to be confident that the outer boundary 
has not affected our results, since radial supersonic outflow is treated exactly in {\sc zeus-2d} \citep{zeus2da}. The total mass loss rate is found by integrating the mass flux across a spherical surface situated at a radius of 85\% of the outer grid boundary (sufficiently far from the outer radial boundary to avoid any effects from spurious reflections due to non-radial outflow); and X-ray `dark' gas is explicitly excluded from our measurement. We then determine the cylindrical radius from within which the integrated mass loss rate occurs by following the streamlines down to the base of the flow. Figure \ref{fig_massl} shows the cumulative mass-loss rate as a function of cylindrical radius, and shows that the mass-loss rate is dominated by the region between  5 and 20~AU, but significant mass loss occurs also at larger radii with 50\% of the mass loss occuring between  18-70AU. This should be compared  to the pure EUV case where only $\sim$10\% of the total mass-loss occurs outside 18AU \citep{font04}.  The X-rays cause a much broader wind profile by being able to heat gas at larger radii than pure EUV irradiation due to their greater penetrative power. The total integrated mass-loss rate over the entire disc was found to be  1.4$\times 10^{-8}M_\odot$yr$^{-1}$.
\section{Inner Hole Flows}
 As will be shown in \S\ref{vis_res}, photoevaporation combined with viscous evolution results in a gap opening in the primordial disc; in contrast to the case of planet clearing, where streams cross the gap at certain azimuths, in the present (axisymmetric) case the inner disc is starved of re-supply once the gap is opened. The inner disc then
drains onto the star leaving the outer disc open to direct irradiation by the central star. The flow expected from such a system is of course different from a primordial disc's photoevaporative flow, and in this section we set out the methods used to calculated mass-loss rates from discs with inner holes.
\subsection{Numerical Method}
\indent Given we expect the mass-loss rates to vary with inner hole size as seen in the EUV case \citep{alexander06a}, we have run several models with inner holes at different radii, out to a radius beyond which disc holes become hard to detect observationally (i.e. at around
$\sim 100$ AU). 
We used a regularly space spherical grid in order to maximise resolution while minimising computational effort. The grid was constructed with 400 cells in the radial direction and 200 cells in the angular direction, again sufficient to resolve the onset of any flow and the scale height of the `dark' disc at all radii. All other {\sc zeus-2d} parameters were kept the same as above. Table \ref{table1} summarises the input parameters used. As initial condition, the  hydrodynamic primordial disc model was cut at the inner hole radius. Then the density profile was  smoothed 
over several pressure scale lengths at inner edge, the we adjusted the angular velocity to match the new imposed radial pressure gradient. The model was then allowed to evolve hydrodynamically until it attained  a steady flow
solution.

\begin{table}
\begin{center}
\caption{Table listing the input parameters for simulations of inner hole discs, where $\rin$ is the inner edge of the X-ray `dark' disc}\label{table1}
\begin{tabular}{c c c}
\hline
Model & $\rin$ (AU) & Grid range (AU) \\
\hline
A & 8.3 & [1.33,90] \\
B & 9.7 & [1.67,90] \\
C & 11.7 & [3.35,100] \\
D & 14.2 & [3.35,100] \\
E & 17.7 & [3.35,100] \\
F & 21.1 & [3.35,100] \\
G & 30.5 & [3.35,100] \\
H & 64.5 & [10,450] \\
I & 86.9 & [15,500] \\
\hline
\end{tabular}
\end{center}
\end{table}
\subsection{Analytic Model}\label{analytic}
 Before we perform detailed hydrodynamic simulations to determine the flow structure of transition discs it is useful to develop a theoretical framework in which to analyze the simulations. In order to estimate the mass-loss rate scaling we adopt a similar method to \citet{alexander06a} and assume the mass loss rate per unit area is $\propto \rho v_l(R)$ where $v_l(R)$ is the local launch velocity of the ionized gas. However unlike the calculations of \citet{alexander06a} where the gas was assumed to be isothermal,  we have no such restriction in our calculations. We expect streamlines
that are launched from the disc's inner rim (which will turn out to dominate the mass loss) to start on
a nearly radial inward trajectory (see Figure \ref{hole_results}). 
Inevitably such a trajectory will result in a centrifugally driven wind (whose force term scales as $1/R^3$) resulting in a launch velocity that scales as the escape velocity $v_e(R)$. Furthermore, if we assume that the effective emitting
area of the wind scales as the square of the inner disc radius ($\rin$) then
we can write the wind mass-loss rate as scaling with $\rho v_e(R) \rin^2 \propto
\rho \rin^{3/2}$. We will use the
code to determine the number density at the base of the flow at the disc's inner rim
(${n_\textrm{\tiny in}}$)  and check out our estimate  that in this case
\begin{equation}\label{mdotin}
\frac{\dot{M}}{n_\textrm{\tiny in}}\propto \rin^{3/2}
\end{equation}
  
The actual scaling of $\dot{M}$ with $\rin$ however needs to be determined numerically
since the ionization parameter at the flow base (and hence $n_\textrm{\tiny in}$)  depends 
on the temperature at the inner rim and this needs to be calculated using the radiation-hydrodynamic algorithm
for each hole size. We parametrize the dependence of mass loss rate on $\rin$
in Section \ref{hole_results}.
\subsection{Results}\label{hole_results}
\begin{figure}
\centering
\includegraphics*[width=\columnwidth]{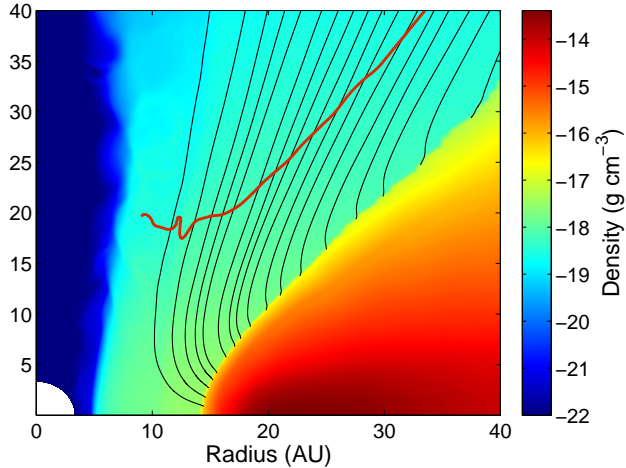}
\caption{The `steady-state' density structure of a transition disc and its photoevaporative wind with an inner hole of 14.2 AU (model D), the streamlines are plotted at 5\% intervals of the integrated mass loss rate, with the sonic surface plotted in red. Note the sonic surface being close to smooth indicates that we are extremly close to a true `steady-state'.  The snapshot shown is at a simulation time of 10 orbital periods at 100AU.}\label{fig_hole}
\end{figure}

All inner hole flows share  the same general morphology: the mass-loss is dominated by regions close to the inner edge, but significant mass-loss still occurs at larger radii. Figure \ref{fig_hole}
 shows the flow structure of model D with an inner hole at 14.2AU. We extract mass-loss rates in an identical fashion to the primordial disc case and note again that the radial structure of the streamlines at large radii allows us to be confident that the outer boundary has not affected our mass-loss rates. 

\begin{figure}
\centering
\includegraphics*[width=\columnwidth]{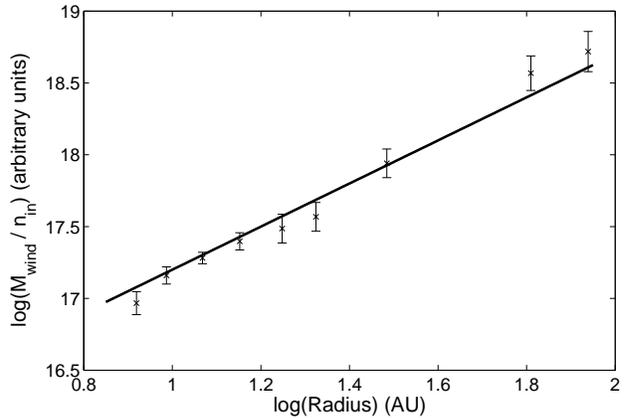}
\caption{Plot showing analytic model ($\dot{M}_{\textrm{wind}}/n_{\textrm{in}}\propto\rin^{3/2}$) agrees well with the simulations. Error bars shown indicate errors in $n_{\textrm{in}}$ obtained by averaging over cells at the inner edge.}\label{check}
\end{figure}

  Figure \ref{check} shows the ratio $\dot{M}/n_\textrm{\tiny{in}}$ reproduces the predicted $\rin^{3/2}$ scaling discussed at the end of \S\ref{analytic}, i.e. it is consistent with our assumption that the effective launch area scales as $R_{in}^2$ and that
the   launch velocity  scales as the escape velocity.

\begin{figure}
\centering
\includegraphics*[width=\columnwidth]{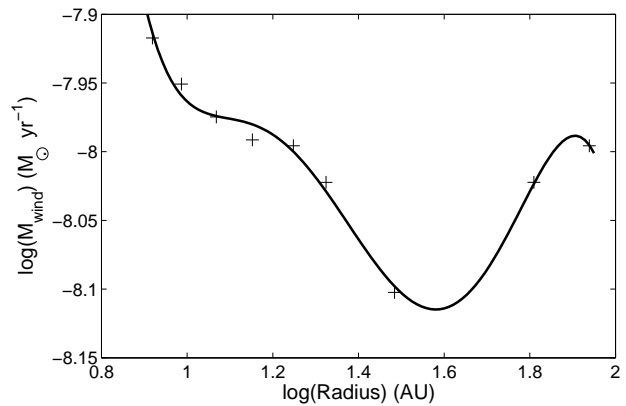}
\caption{Plot showing $\dot{M}$ at different inner hole sizes; The blue point represent mass-loss rates from models A-I, while the black line shows the numerical fit used in the viscous evolution.}\label{mdot-hole}
\end{figure}

In Figure \ref{mdot-hole}
 we present the mass-loss rates as a function of inner hole radius complete with the numerical fit used in the viscous evolution. This shows that for $\rin$ out to $\sim 30-40$AU the mass-loss rates fall approximately as $\rin^{-0.3}$ while at larger $\rin$ the mass-loss rates begin to increase. 
We can understand this behavior by noting that equation (\ref{mdotin})
and the definition of the ionization parameter implies that 
$\dot M \propto \xi^{-1} \rin^{-0.5}$ so that the scaling of photoevaporation
rate with radius depends on the radial variation of $\xi$ and hence
on the form of the temperature versus  ionization parameter relation. From
Figure \ref{fig:relationsb} we see that this contains a point of inflection
at $\log \xi = -5.2$, $\log T=3.2$ which is encountered when the $\rin$ is
 around $40$ AU. At smaller radii and hence higher temperatures the variation in ionization parameter is small for a given temperature variation and hence the 
modest decline in the flow temperature with inner hole radius is achieved at nearly
constant $\xi$. Beyond $R_{in}=40$ AU, a modest decline in temperature
corresponds to a larger decline in $\xi$  and hence the fall off in
$\dot M$ becomes less steep and even starts to increase somewhat with
increasing $\rin$. It should however be stressed that the radial variation
of $\dot M$ with $\rin$ is very mild  (see Figure (\ref{mdot-hole})). 

\section{Tests}\label{test}
\begin{figure}
\includegraphics[width=\columnwidth]{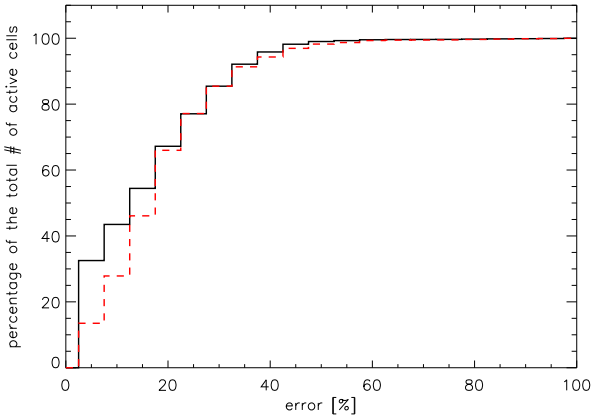}
\caption{Histogram of cells with temperature errors less than a given value. Black solid line refers to
errors in cells in entire grid, red dashed line to  errors from cells that are locally sub-sonic (in the rotating frame).}\label{compare}
\end{figure}

\begin{figure}
\includegraphics[width=\columnwidth]{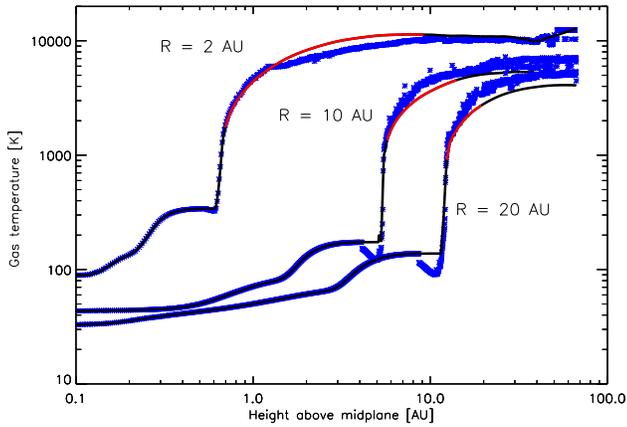}
\caption{Radial cuts showing temperature comparison for R=2,10,20AU. Solid line indicates temperatures calculated by the {\sc zeus-2d} algorithm, where the red solid line indicates where those cells are locally subsonic (in the rotating frame) and in the photoevaporative flow. Blue stars represent the temperatures determined by {\sc mocassin}.}\label{cuts}
\end{figure}

\begin{figure}
\centering
\includegraphics*[width=\columnwidth]{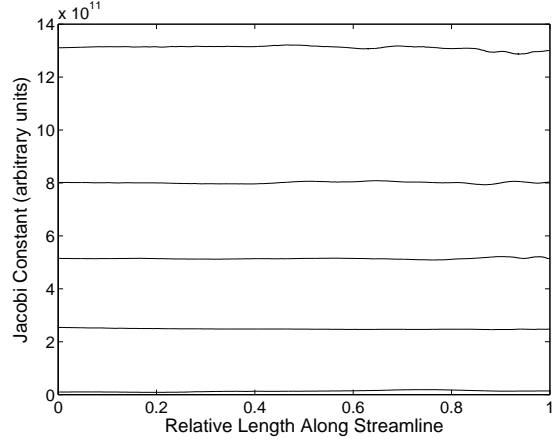}
\caption{Plot of the Jacobi potential along various streamlines in the range 2-50AU, plotted against the normalised streamline length. The variation in the mean value along each streamline is $\sim$ 3\%}\label{jacobi}
\end{figure}

In order to be confident that the solutions obtained are accurate and are unaffected by the assumptions discussed in \S\ref{hydro}, we have conducted various tests which we discuss below: 
\begin{enumerate}
\item A numerical convergence test using double the resolution confirms the flow is converged to $\sim$5\%. 
\item In order to be confident the parametrization of X-ray heating is correctly reproducing the temperature structure in the flow we have re-calculated the gas temperatures of the steady-state solution using {\sc mocassin}. Figure \ref{compare} shows a comparison between the temperatures determined by the {\sc zeus-2d} algorithm and those obtained using {\sc mocassin}. This shows that we have errors $<$15\% for 50\% of all grid cells and errors $<$40\% for 95\% of the grid. In Figure \ref{cuts} we show the comparison between the gas temperatures determined using the {\sc zeus-2d} algorithm and those obtained using {\sc mocassin} at various cuts in cylindrical radii. It shows that we have good agreement throughout the grid especially in the launch region of the flow (region of large temperature gradient) and in the regions of the flow which are subsonic (red regions) where the thermal pressure gradient can be a first order effect. We note that the disagreement between the temperatures in the flow/disc transition region is an artifact of {\sc mocassin} which imposes a hard transition between the molecular and atomic zones (see discussion in ECD09), hence the errors quoted can be considered an overestimate.
\item It has been assumed that the gas and dust are thermodynamically coupled in regions that are optically thick to X-ray irradiation. In order to test this assumption we have performed a run where we allow the gas in the  X-ray dark regions to evolve adiabatically. We then compare the timescale for the gas and dust  to return to thermal equilibrium through collisions \citep{whitworth97} to the local dynamical timescale and we find that the dynamical timescale is $\sim 100$ the thermal timescale at the minimum. Thus we can conclude that the gas and dust will be thermodynamically coupled when not penetrated by the X-ray flux.
\item In the following sections, we assume that the mass-loss profile obtained for the primordial disc is valid for the majority of the discs lifetime (i.e. the viscously accreting phase). During this period the disc structure will change as material drains onto the central star. In order to check our mass-loss rate is is valid for a varying (optically thick) disc structure, we compare it to the mass-loss rates obtained for the adiabatic runs mention above, which produce markedly different (adiabatic) disc structures and find good agreement between the models. 
\item In our hydrodynamic modeling we have used the dust temperatures (and hence gas temperature of the X-ray `dark' region) obtained by \citet{d'alessio01}. Since the dust temperatures will be linked to the disc structure and vice-versa, and since the disc structure has evolved from the original hydrostatic equilibrium structure, is important to assess how accurately the dust temperatures of the \citet{d'alessio01} model can describe the dust temperatures in the hydrodynamic solution. We've calculated the dust temperatures of both the original model and the flow solution using {\sc mocassin} and find we have agreement to within the Monte-Carlo errors.
\item In order to determine whether our models are close to a true steady-state we compute the Jacobi\footnote{The Jacobi potential is the analogue to the Bernoulli potential in a rotating frame.} potential along the streamlines, noting that this will be constant in a steady flow. For each streamline we determine the enthalpy term by paramatrizing the run of pressure and density as a baratropic relation (i.e. $P(\rho)$) along the streamline, then integrating along the streamline. However we note that, the flow is not strictly baratropic or irrotational and thus don't expect the Jacobi potential to be constant for all streamlines. Figure \ref{jacobi} shows that the Jacobi constant is conserved along the streamlines. We find that the Jacobi constant is conserved to within 3-4\% indicating we are very close to a true steady-state flow.
\item  ECD09 models do not include molecular cooling in the thermal
balance. While (as shown in Figure \ref{fig:flowa}) the flow is launched in a
region higher than the optically thick disc, the lack of molecular
cooling could in principle be a potential source of error on the gas
temperature in the disc interiors, and as such the cause of a
hydrostatically `puffed up' disc. Although as shown by \citet[][their
    Figure 4]{glassgold04} molecular cooling only comes in at large
column densities and, even there, it only has a very limited role when
compared to dust-gas collisions, it is useful to test whether a cooler 
(flatter) disc interior would result in significant changes in the
photoevaporative mass loss rate of the system (due to neglecting molecular cooling, or some other, as yet unidentified cooling mechanism).  
We can do this by
considering the mass-loss rates obtained from the adiabatic models
described above (point iii \& iv). The use of an adiabatic
equation of state results in a cooler disc structure
and hence a lower launching height. While the exact
quantitative result is sensitive to disc heating, the
overall mass-loss rates in all models were found to
be $\sim10^{-8}$ M$_\odot$ yr$^{-1}$. This effect is small since the
change in gravitational energy between the midplane
and the launching height is of order 10\% of the total gravitational energy gained by the wind, as shown in Figure \ref{energy}. 
\end{enumerate}


\section{Viscous Evolution}\label{vis_setup}
The evolution and dispersal of an irradiated protoplanetary disc is controlled by the simultaneous actions of the photoevaporative wind and viscous torques in the disc. A numerical simulation of disc dispersal must therefore simultaneously account for photoevaporation and viscous evolution. 
\subsection{Method}
The diffusion equation for the evolution of the surface density is \citep{pringle74,pringle81,alexander06b}:
\begin{equation}\label{diff_eqn}
\frac{\partial\Sigma}{\partial t}=\frac{3}{R}\frac{\partial}{\partial R}\left[ R^{\frac{1}{2}}\frac{\partial}{\partial R}\left(\nu(R)\,\Sigma\, R^{\frac{1}{2}}\right)\right]-\dot{\Sigma}_w(R,t)
\end{equation}
Given that the initial disc structure taken from \citet{d'alessio98} has a surface density that scales approximately as $\Sigma \propto R^{-1}$, an initial surface density structure of the form
\begin{equation}\label{eqn:disc}
\Sigma(R,0)=\frac{M_d(0)}{2\pi R_1 R}\exp(-R/R_1)
\end{equation}
is used, where $M_d(0)$ is the initial disc mass and $R_1$ is the initial scale size of the disc. This initial structure is a similarity solution of the viscous diffusion equation \citep{pringle74} when the viscosity scales linearly with radius and can be parametrized as
\begin{equation}
\nu=\nu_1R/R_1
\end{equation}
and the mass-loss in the wind can be neglected. The viscous scaling constant $\nu_1$ can be determined from \citep[see][]{hartmann98,clarke01,alexander06b}:
\begin{equation}
\dot{M}(t=0)=\frac{3M_d(0)\nu_1}{2R_1^2}
\end{equation}
The integration of equation (\ref{diff_eqn}) is performed using a first-order explicit scheme, the grid consists of 1000 points equispaced in $R^{\frac{1}{2}}$ and spans the range [0.0025, 2500] AU with  zero-torque boundary conditions \citep{pringle86}. The radius beyond  which $e^{-1}$ of the  disc mass is initially located  ($>R_1$) is taken to be 10AU and the initial accretion rate is 
$5 \times 10^{-7} M_\odot$ yr$^{-1}$.  While we adopt a viscosity scaling $\nu\propto R$ in order to be consistent with the $R^{-1}$ surface density profile of our initial disc structure, \citet{clarke01} \& \citet{alexander06b} considered the effect of using a different viscosity parametrization and concluded that, while the absolute timings would change, the overall qualitative nature of a photoevaporative `switch' was unchanged. 

The relative timescales between the three stages (viscously evolving, gap opening and disc clearing) are independent of the initial disc mass (for a fixed initial accretion rate) and can thus be scaled accordingly for any chosen value. In order to set a reasonable absolute value
for the disc lifetime, we have chosen an initial disc mass of 0.2M$_\odot$, which (given the
initial accretion rate) corresponds to an initial viscous timescale at the disc's outer
edge of $2\times10^5$yr; this in turn corresponds, for our chosen case of $R_1=10$ A.U. to a viscous
alpha parameter of $\sim 10^{-2}$ at $R_1$.
\subsubsection{Surface Mass-loss rates}
The functional  forms of $\dot{\Sigma}_w$ used are those obtained for the primordial disc models and inner hole models. Since we expect the inner hole to drain on a short timescale, as a first approximation we either use the primordial disc wind profile, or the inner hole mass-loss profile and do not attempt to model the transition in any detail. The transition occurs once the radial column density to the inner edge is $<10^{22}$cm$^{-2}$.
\subsection{Results}\label{vis_res}
\begin{figure}
\includegraphics*[width=\columnwidth]{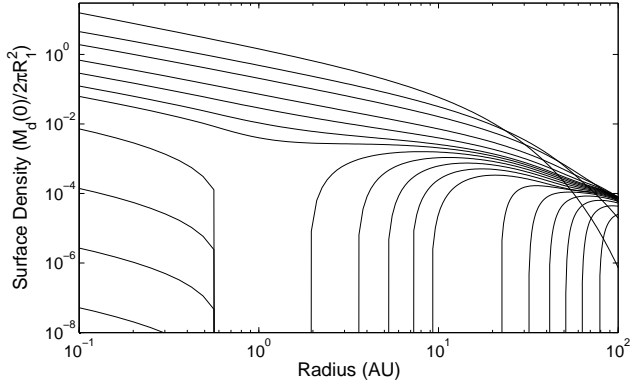}
\caption{Snapshots of the surface density plotted at 0\%,~20\%,~40\%,~60\%,~70\%,~75\%,~77\%,~79\%\ldots99\% of the disc's total lifetime (see footnote 4), The surface density is in dimensionless form and in units of $M_d(0)/2 \pi R_1^2$ (see equation \ref{eqn:disc}). }\label{vis_evolfig}
\end{figure}

Figure \ref{vis_evolfig} shows the evolution of the surface density of the disc. At early times the evolution of the disc is similar to a standard viscous accretion disc with a power-law decline in surface density and accretion rate. Once the accretion rate is comparable to the mass-loss rate due to the wind (which occurs at $\sim 60 \%$ of the disc lifetime{\footnote {At this point we define the disc lifetime as being the total lifetime when photoevaporative clearing is taken into account; i.e. the time to clear out to 87AU (beyond which $\dot{M}$ is not accurately known); Due to the power law nature of viscous draining, it is hard to define a `lifetime' of a control disc (with no photoevaporation). However, to be indicative: the model disc that we have run has a total lifetime when photoevaporation is included of $\approx 1.3$ Myr; if such a disc were simply viscously evolving without photoevaporation it would become optically thin in the region of the disc that dominates the K band emission after $\OO (100)$ Myr.}}), the photoevaporative wind begins to affect the surface density profile in the range 5-30 AU;  a gap then opens at  1AU after  78\% of the disc lifetime. Once the gap opens,  the inner disc quickly drains on a timescale of order the viscous timescale at  1  AU. During this draining time the inner edge of the outer disc is continually photoevaporated outward due to the significant mass-loss at larger radius.

\indent Once the inner disc has drained, the outer disc is exposed to the direct photoevaporative wind, and the inner edge of the disc has now moved to  9AU. The total mass remaining in the disc after the inner disc has drained onto the star was  9\% of the intial disc mass and the outer disc is then dispersed entirely in the remaining  15\% of the disc's lifetime.  We find, in total 84\% of the intial disc mass is accreted onto the central star, with the remaining being lost in the wind.

\begin{figure}
\centering
\includegraphics*[width=\columnwidth]{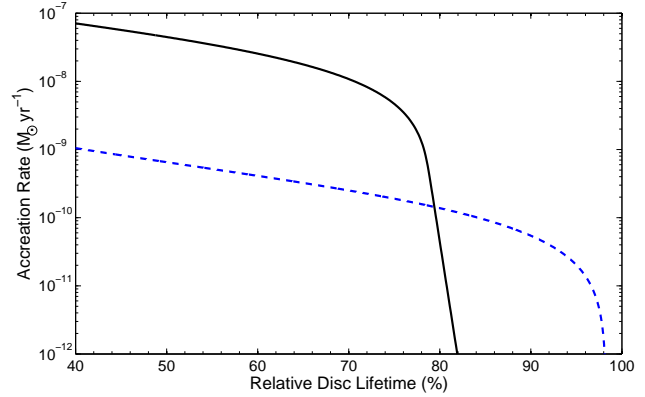}
\caption{Plot of accretion rate onto the star against time. The black solid line shows the model for the X-EUV wind, while the blue dashed line shows the pure EUV canonical model \citep[see][Figure 1]{alexander06b}. }\label{mdot_time}
\end{figure}

In Figure \ref{mdot_time} 
we show the accretion rate onto the star throughout the disc's lifetime and   show that the disc evolves much faster compared to a standard viscously accreting disc. We also see  that, even prior to gap opening, 
the disc spends a significant portion of its lifetime ($\sim 15\%$) with  the
accretion rate onto the star being  less than  the photoevaporative wind rate: the accretion rate onto the star is   1$\times10^{-9} M_\odot$ yr$^{-1}$  at the point that  the gap opens, which is about an order of magnitude less than the wind mass loss rate. This
effect (which has been termed `photoevaporation starved accretion' by \citealt{drake09}) is a consequence of the radially extended
mass loss in X-ray heated discs so that, prior to gap opening, there is a significant region of the disc where the accretion
rate decreases towards small radii. This behavior was  not seen to be a prenounced effect in  previous pure EUV models since,
in the EUV case, the mass loss is more strongly localised  and the gap opens in a disc that is hardly depleted by the wind
at large radii.

\section{Comparison with previous work}
Several other authors have considered photoevaporative flow driven by either, EUV or X-ray radiation and \citet{ercolano09} have considered the combined effect of X-EUV radiation. The predicted mass loss rates due to pure EUV photoevaporation have been accurately determined as $1.1\times 10^{-10}$M$_\odot$yr$^{-1}$ \citep{font04} for a 0.7M$_\odot$ star and an ionizing flux of $10^{41}$s$^{-1}$. Also \citet{alexander06a} found that when the inner hole first forms, the mass loss rate rises by about a factor ten and then increases mildly as the hole grows according to  $\rin^{1/2}$. 

\indent As discussed in Section 1, previous  estimates of  X-ray photoevaporation rates to be found in the literature are qualitatively different from the values presented in the present work since they are not  based on hydrodynamic calculations. 
\citet{alexander04} estimated   mass-loss rates per unit area at $\sim 20$ AU which were  similar to those obtained in  this work. ECD09 argued for  similar surface mass-loss rates   as those of \citet{alexander04}, but estimated a much larger launching regions (i.e.over a region 5-50AU). They therefore  concluded that X-ray photoevaporation would dominate over EUV photoevaporation at all but the lowest of X-ray luminosities. The estimates of \citet{gorti09} suggested   much lower mass loss rates than the EUV case, implying  that X-rays would only have an indirect effect on disc dispersal. ECD09 include a detailed discussion of the differences in the thermochemical calculations of Gorti and Hollenbach 2009, which may have led to their result, and concluded that the rather harder X-ray spectrum employed by Gorti and Hollenbach may have been an important contributory factor.  

\indent The mass loss rate of  1.4$\times 10^{-8}$M$_\odot$yr$^{-1}$ that we obtain from our radiation hydrodynamical calculations can be compared to the estimate of ECD09 for the same input parameters (model FS0H2Lx1 of ECD09) of 6.7$\times 10^{-9}$ M$_\odot$yr$^{-1}$. This difference arises from the fact that ECD09 assumed that the gas escaped at the sound speed at the point at which the internal energy of the gas was equivalent to the escape potential. In reality the flow is launched sub-sonically from deeper in the disc where the density is higher. It is nevertheless  reassuring that the two values are of a similar order of magnitude. We also find that the mass-loss  occurs over a large radial range of 1-70AU, in contrast to  the EUV case where it is  steeply peaked about  one characteristic radius (for a 0.7M$_\odot$ star this would be 6.2AU). Given the mass loss rates obtained and the monotonic temperature : ionization parameter relations found (see Figure \ref{fig:relations}),  it is clear in this case that X-ray photoevaporation will be a dominant effect in disc evolution.  

\indent Coupling the mass-loss profiles to secular viscous evolutions models has only been previously attempted for the pure EUV case by \citet{alexander06b} where the mass-loss profile had been accurately determined.  In order to compare our model with what the
EUV model of \citet{alexander06b} would predict with the same input parameters, we take the canonical model
\citep[see][Figure 1]{alexander06b} and multiply both the initial disc mass and the initial disc viscous timescale by a factor $4$, thus
ensuring that the rescaled model has the same initial accretion rate, disc mass and viscous timescale as the
present model. \footnote{Ignorance of the magnitude of the viscosity in discs means that the absolute values for disc
lifetimes that we quote here in the EUV and X-ray heated case should not be used as grounds for favoring one model
or another: we have scaled the model used here so that its total lifetime is an observationally reasonable $\sim$1Myr and the EUV lifetime
is then rather long, but we could have just as easily scaled down the initial mass  and initial viscous timescale in such
a way that the EUV lifetime was observationally reasonable and the X-ray lifetime was unreasonably short. The purpose of our
present comparison is therefore to look at the {\it ratio} of timescales in X-ray and EUV models with the same underlying
viscous parameters.} In the (rescaled) EUV case, it  was found that the disc dispersed entirely after $\sim$24Myr
with the transition stage occurring on a short $\sim10^5$yr timescale (approximately 3\% of the total disc lifetime).  
Our results are consistent with this `switch' nature of disc dispersal (albeit in
somewhat softened form: see Figure \ref{mdot_time}). In the present (X-EUV) models we find that  a  gap in the disc opens
at $\sim$1AU after  78\% of the discs lifetime (i.e. nominally at an age of  $\sim$1Myr). Once the inner edge of the outer disc is exposed to direct radiation we find the outer disc is quickly removed on a timescale  $\sim10^5$yr with a mass loss rate scaling as $\rin^{-0.3}$.  

\begin{figure}
\centering
\includegraphics*[width=\columnwidth]{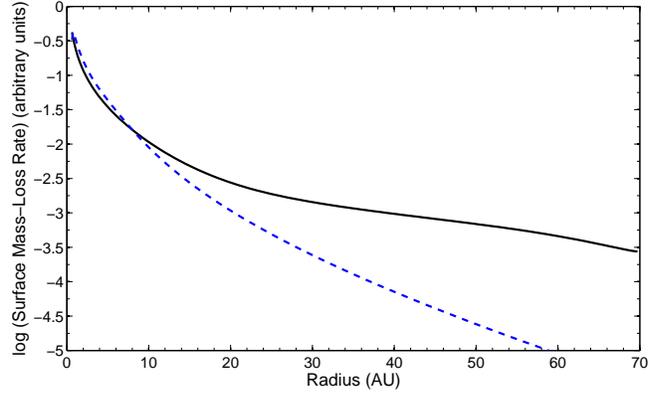}
\caption{In this Figure we plot the surface mass-loss rate for the X-EUV wind (solid black line) obtained in this work and the EUV wind (dashed blue line) \citep{font04,alexander08}. We have scaled the peaks of the profile to be the same value in order to compare the relative broadness of the different profiles. Clearly this shows the X-EUV profile is much broader than the EUV profile.}
\label{windcompare}
\end{figure}

 While X-ray driven photoevaporation produces the same qualitative results when it comes to disc dispersal, there are several distinct differences. Prior to gap opening the surface density begins to flatten, indicating that the accretion rate is dropping below that of a steady disc (i.e. that the disc is in a state of `photoevaporation starved accretion' as discussed above). 
This  effect is much less obvious in the EUV case owing to the more restricted  radial extent of the
EUV wind. 
Once the gap opens at  around a few AU (similarly  in both cases), the inner edge of the outer disc evolves rapidly to $\sim 10$AU in the X-ray case due to the fact that this region was already significantly depleted, due to the broad mass-loss profile. In  the EUV model,  the inner edge remains at 1AU during disc draining, due to the steep mass-loss profile and then  quickly moves outwards to $\sim 20$AU  as the photoevaporation rate  rises by an order of magnitude
when the inner edge of the disc is directly irradiated. The difference in behavior of the edge of outer disc during inner disc draining, can be readily understood when the surface mass-loss profiles are compared for the EUV case and X-EUV case. Figure \ref{windcompare} clearly shows the X-EUV profile is much broader than the EUV profile; the EUV wind thus
has very little effect on the disc beyond $r_g$ whereas the X-EUV wind subjects the disc to a period of depleted accretion inwards of 25AU prior to gap opening.
    During the clearing of the outer disc, the mass-loss rate falls as $\rin^ {-0.3}$ (out to 30-40AU) in the X-ray case, compared to increasing as $\rin^{1/2}$ in the EUV case.
This difference is due to the fact that  
in the EUV case, where the ionized region is approximately isothermal, the flow is launched at roughly
constant ($\sim$ sonic) velocity and a simple Str\oe mgren sphere argument implies that the density in the launch
zone scales as $\rin^{-1.5}$;  in the X-ray case, where a range of temperatures are present, the effective launch speed from the exposed inner rim of the disc is the local escape speed which scales
as $\rin^{-1/2}$. The density in the launch zone (which corresponds to a nearly constant value of the ionization 
parameter) falls off slightly less steeply than $\rin^{-2}$. Both these effects mean that EUV photoevaporation is
relatively favored at large radius.  We estimate that the radius at which the direct EUV photoevaporation rate becomes comparable to the X-ray photoevaporation rate is $>200$AU and hence the EUV will  remain  unimportant in the case of discs around solar type stars with X-ray luminosities $\sim$10$^{30}$erg s$^{-1}$ and greater. It may however   be relevant for systems with lower X-ray luminosities. Given the large scatter of X-ray luminosities for stars of similar masses, a full discussion of the parameter space is beyond the scope of this paper, but will be considered in future work.    

\section{Model limitations}
While this model is an important step forward in understanding disc dispersal, it is important to discuss the limitations of our models. We have neglected the effects of non-ionizing FUV radiation in our calculations. \citet{gorti09} have shown that FUV radiation is capable of heating gas to above the escape temperature at large radius (100-200AU), and predict mass loss rates of a similar magnitude to those obtained from our X-EUV models. Given that these mass-loss rates are far more uncertain than those obtained from a self-consistent hydrodynamic model we cannot make any definite predictions of the effect FUV inclusion would have on our model. Qualitatively, however, the existence of a further FUV wind at large radius would starve the inner disc of material and would thus accelerate the final
inside-out clearing described here.

\begin{figure}
\centering
\includegraphics*[width=\columnwidth]{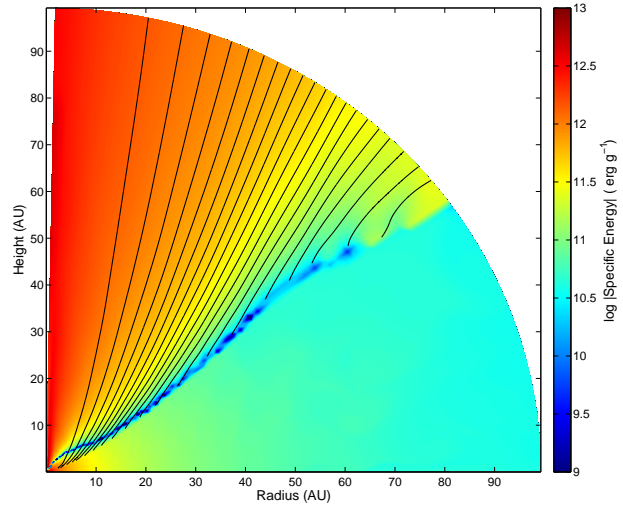}
\caption{Figure showing the specific energy of gas particles throughout the grid. Streamlines are plotted at 5\% of the total mass-loss rate. The dark blue lane indicates the transition between a overall negative specific energy (bound gas) and an overall positive specific energy (unbound gas). This clearly shows that the total energy gained by gas launched into the wind is less than $L_X$ with a total efficiency $\sim 8\%$ compared to the input X-ray energy.}\label{energy}
\end{figure}

Throughout our modeling we assume that the gas is always in thermal equilibrium with the radiation based on the comparisons between the thermal and flow timescales discussed in \S\ref{mocassin}. This assumption can be tested by comparing the gained energy flux  of the wind to the ionizing luminosity.  Figure \ref{energy} shows a plot of specific energy for the primordial disc. Considering the streamlines around the median mass-loss value (launching from 15-20AU)  the total change in specific energy along these streamlines from launch (with a \emph{negative} specific energy of $\sim10^{11}$ erg g$^{-1}$) to exit from the grid (with a \emph{positive} specific energy of $\sim 5\times10^{11}$ erg g$^{-1}$) is $\sim 6\times10^{11}$ erg g$^{-1}$.    Given the grid boundary is far from the sonic surface (see Figure \ref{fig:flowa}) this can be considered close to the maximum gained energy flux of the wind (c.f. Parker wind; \citealt{parker58}). In order to compare with the input energy rate ($L_{tot}$), we can compute the mechanical luminosity of the wind using $L_{mech}\approx\dot{M}_w\times \Delta e$ (where $e$ is the specific energy) which evaluates to $\sim 10^{29}$ erg s$^{-1}$. Thus it is clear that the energy gained in the wind is much less than the input energy from the X-EUV spectrum ($4\times 10^{30}$ erg s$^{-1}$).  For all flows (both the primordial and inner-hole flows) we explicitly calculate the  gained energy flux in the wind and find that its $\ll L_\textrm{\tiny X}$ ($\sim <$8\%). Hence the assumption of thermal equilibrium is valid in this case. Hence the terms in the energy equation from advection and $p\dd V$ work can be neglected.

\indent During the transition between a  primordial disc and a transition disc, we assume the mass-loss profile for the former, until the column to the inner edge of the outer disc is less than $10^{22}$cm$^{-2}$. In reality we would expect the transition to be a smooth process. Given the difference in mass-loss rates between the primordial and transition disc models is small, and the transition time is rapid, such an approximation will have little effect on the overall evolution. However a more detailed approach will be required when it comes to predicting the statistical distribution of inner hole sizes as a function of stellar parameters \citep{alexander06b}.
     
\section{Observational Consequences}
We have shown that the X-EUV photoevaporation model described here  can reproduce the `two-timescale' evolution  observed in young solar type stars, in agreement with current observational constraints on relative timescales. In this sense it is
qualitatively similar to the EUV switch model as set out in \citet{alexander06b}: in particular  the second (rapid clearing) timescale is similar
and so is the radius ($\sim  1$ AU) at which the gap first opens. The temperature distribution in the flow is however rather different in the two cases (the X-EUV wind is somewhat cooler: compare Figure \ref{cuts}  with the EUV case where the wind is close to $10^4$K throughout). It is thus an immediate task to
compare the predictions of the new model with observational diagnostics that have been studied in the EUV case (Ercolano et al in prep.).  In particular,  the strength and line profile of the  [Ne~{\sc II}] emission at $12.8 \mu$m \citep{pascucci07,pascucci09,herczeg07,najita09,flaccomio09} has been suggested as an observational test for X-ray irradiated discs e.g.\citep{glassgold07,meijerink08,guedel09,schisano09}; our calculations open up the possibility of detailed modeling, analogous to that of \citet{alexander08} for   the EUV only case. Likewise, the X-EUV model needs to be assessed with regard to the
production of forbidden line emission (including O{\sc[I]}) as in the study of \citet{font04} for the EUV case.

 The important difference in terms of disc secular evolution between the X-EUV and the EUV only
model is  the {\it magnitude}  of the photoevaporative mass loss rate which is larger 
in the X-EUV case  by  2 orders of magnitude. The
wind mass loss rate is now  comparable with  typical accretion rates
measured in Classical T Tauri stars \citep{natta06}. This means that whereas the EUV  
switch provided a mechanism for the final shutdown of a protostellar accretion disc when its accretion rate had fallen to very low (nearly undetectable) values, the X-EUV wind (at least for the rather high X-ray luminosity employed 
here) should cut off a disc in its prime.  This raises a number of issues:

\subsection {Accretion rates in T Tauri stars and their relationship with X-ray luminosity}

 If all discs were subject to the vigorous X-EUV winds found in this model, then young stars
should rarely exhibit accretion rates less than a few times $10^{-8} M_\odot$ yr$^{-1}$, since
subsequent draining is rapid (see Figure \ref{mdot_time}). This is clearly untrue observationally, as this accretion rate
is close to the median for Classical T Tauri stars. However it needs to be borne in mind
that T Tauri stars exhibit a $\sim 2$ orders of magnitude scatter in
in  X-ray luminosities at a given mass \citep{preibisch} and that the
value that we have adopted here is modestly above the median value.  
We have not yet undertaken full hydrodynamical calculations with a range of X-ray luminosities but the hydrostatic estimates of ECD09 tentatively  suggest that the photoevaporation rate may scale roughly linearly with X-ray luminosity
(cf the EUV case which, being recombination limited, predicts a mass loss rate that scales as the
square root of the EUV output of the star; \citealt{hollenbach94}). 
Therefore we would expect that
in stars with a lower X-ray luminosity the  plot of
accretion rate versus time (equivalent to Figure \ref{mdot_time})
would steepen at a lower accretion rate. 

 What is the expected sign of the dependence of accretion rate upon
X-ray luminosity within the subset of stars that have not yet cleared
their discs (i.e. in Classical T Tauri stars)? 
Some care is needed
in phrasing this question. If we have a coeval cohort of stars, then evidently
the stars that are stronger X-ray emitters would be expected to exhibit
lower accretion rates, as a consequence of `photoevaporation starved
accretion'. On the other hand, if we have a mixture of (disc bearing)
stars of all ages
then in a statistical sense the strong X-ray emitters will be younger
(because the predicted disc lifetime is less) and they will have higher
accretion rates on average (because the low luminosity X-ray sources
live for a longer time and spend most of this time with accretion rates
modestly above the (low) wind rate). 

 A clear observational consequence of the lower disc lifetime in luminous
X-ray sources is that such sources spend a higher fraction of the pre-main
sequence state in a disc-less (Weak Line T Tauri) state: consequently
one expects that Weak Line T Tauri stars should be  more luminous
X-ray sources on average
than their disc bearing (Classical T Tauri) counterparts. This is indeed
the case  \citep[e.g.][]{neuhauser95,flaccomio03,preibisch} though this  has also been variously explained in terms either
of X-ray absorption in accretion columns \citep{gregory07} or confinement
of the X-ray producing corona in accreting systems \citep{preibisch} (in other words,
it is conventionally assumed that accretion suppresses X-ray emission
rather than the effect predicted here whereby X-ray emission suppresses
accretion: see also \citealt{drake09} for a discussion of this possibility).
Naturally, this effect cannot be quantified until we have verified through
hydrodynamic simulations how the wind mass loss rate scales with
X-ray luminosity. 

\subsection{Disc lifetimes}
 The fact that the X-EUV photoevaporation rate is much higher than that for EUV only has obvious implications for disc lifetimes. For example, we find that for the same initial parameters, the
disc lifetime is around 1.5 orders of magnitude lower in the present case than when compared with
the EUV only calculations of \citet{alexander06b}. However it should be borne in mind that in both cases the
absolute lifetime depends on the the time for viscous evolution down to a critical accretion rate and this therefore depends both on the initial disc mass and the viscosity in the disc. The magnitude of the latter can in  any case be adjusted so as to reproduce any reasonable disc lifetime. What will be interesting, however, is to examine the {\it distribution} of relative disc lifetimes that results for a plausible distribution of disc initial properties and X-ray luminosities \citep[c.f.][]{armitage03}.

\subsection{ Transition discs (inner hole sources)}
  
  We have shown that X-ray luminous T Tauri stars should evolve through
the following  sequence of  states: a) optically thick (untruncated)
stage dominated by viscous evolution b) inner gap phase, with a low
density inner disc and accretion on to the central star that is less
than the wind mass loss rate ( $< 10^{-8}
M_\odot$ yr$^{-1}$) and c) inner hole phase with an evacuated inner disc
and essentially no accretion onto the star and an outer disc mass of
$< 0.01 M_\odot$. This is qualitatively similar to the EUV case described
in \citet{alexander06b}; important quantitative differences are the
($\sim 100 \times$) higher upper limit on the accretion rate in stage
b) in the X-EUV case and also the somewhat higher upper limit (roughly
an order of magnitude) on the outer disc mass during stage c). 
It has often been argued that the relatively high accretion rates
and outer disc masses in observed inner hole sources is a counter-argument
against the relevance of (EUV) photoevaporation \citep{najita07,Rreview,kim09}. We thus see that X-EUV photoevaporation
alleviates both of these problems.

 It would however be unwarranted to necessarily conclude that all
(or even the majority) of observed T Tauri stars with spectral
evidence for cleared inner holes  are created in this way.
Indeed it should be stressed that even if we judge that the bulk of such
sources are created by other mechanisms, this does not diminish the
important role of photoevaporative  clearing in the majority of discs. 
It needs to be recalled that 
the evolutionary sequence described above implies - in the
absence of other intervening processes (such as the formation of
planets or grain growth) - that {\it all} disc systems should evolve through
a {\it short-lived} inner hole phase. The observed frequency of 
inner hole sources (of order $10 \% $ of all disc-bearing sources)
is broadly compatible  {\it either} with 
a scenario in which all observed inner hole sources are a short-lived
evolutionary phase undergone by all stars {\it or} one in which the
majority of observed inner hole sources are instead a long-lived evolutionary
phase undergone by a subset of all T Tauri stars. There is no
shortage of other mechanisms proposed for inner hole creation, including
for example clearing by planets or binary companions \citep{quil04,van06,rice03,rice06}, grain growth
in the inner disc \citep{dullemond05,ciesla07}, photophoresis \citep{krauss07} and inside out evacuation by the magnetoroational
instability \citep{chiang07} (see e.g. reviews
by \citealt{Rreview}  and discussion in \citealt{kim09}); what is
notable is that  these alternative mechanisms are not linked
to a mechanism for clearing the residual outer disc and  are thus not
associated with 
a short-lived phase prior to complete disc dispersal. 
Therefore although they may apply to some of the observed
inner hole sources they cannot represent an evolutionary norm for
young stars without over-predicting the incidence of inner hole sources.

  Given that we cannot distinguish on statistical grounds between a short-lived transistion
phase experienced by all stars or a long-lived phase undergone by a minority,
we need instead to look at the properties of individual sources, bearing
in mind that the properties of these sources will partly depend on the
observational criteria used for object selection. Therefore, for example,
the sample of inner hole sources compiled by \citet{kim09} on the
basis of a dip in the Spitzer IRS spectrum (in the range $5-40 \mu$m)
are largely a group of objects with high accretion rates on to
the central star ($\sim 10^{-8} M_\odot$ yr$^{-1}$) and high outer disc
masses ($\sim 0.01 M_\odot$). Whereas the X-EUV models are close to being
able to reproduce these conditions in a short period briefly after gap
opening, they should not be representative of photoevaporative holes
in general. In particular, X-EUV models are not compatible with large
($> 20 $ A.U.) holes that are vigorously accreting (such as SZ Cha, GM Auriga
or UX TauA; \citealt{najita07}, \citealt{calvet05}, \citealt{espaillat07}).
On the other hand, the inner hole sources compiled by \citet{cieza08}
are qualitatively different, being associated with lower disc masses
and low or undetectable levels of accretion: these may indeed include
sources with photoevaporating holes in the evolutionary stage c) above.
(Likewise the mm survey of Andrews \citealt{andrews05} contains objects
that may be in this category, although none of these have the high
$\sim 0.01 M_\odot$ disc masses that are predicted by X-EUV models
at the onset of outer disc clearing). 
It should be stressed that EUV photoevaporation models are mildly biased
towards larger hole sizes and that this effect is somewhat stronger in
X-EUV models, due to the larger outer disc masses and the
gentle decline in wind rate at large radii (Figure \ref{windcompare}). We
would therefore expect that photoevaporating holes would be more strongly
represented amongst sources with large inner holes as one would expect
to uncover with Herschel or ALMA. 

  A final point about the formation mechanism for  observed inner hole  
sources is that it is a robust prediction of photoevaporation models
that there should be no flow across the gap once it is opened. Thus
following gap opening, the  draining inner disc is not re-supplied and
the ratio of its mass to the accretion rate onto the star should 
be a good measure of its lifetime in the draining phase. Observed
inner hole sources are generally characterised by the dust content
of their inner discs and in most cases the gas mass is unknown.
However, measurements of CO emission in TW Hydra and GM Auriga
\citep{salyk07} indicate very low inner disc gas masses, which,
when combined with the measureed accretion rates in these objects would
imply improbably short lifetimes in the absence of re-supply. 
This argument therefore is a strong indication that the
inner disc is re-supplied in these objects, as is a natural expectation
in planet clearing models.

\section{Conclusions}
In this work we have presented the first self-consistent hydrodynamic model of an X-EUV driven photoevaporation wind. The mass-loss rate of   1.4$\times 10^{-8}$M$_\odot$yr$^{-1}$  occurs over a large radial extent of  1-70AU. This should be contrasted with the much lower $\sim 10^{-10} M_\odot$ yr$^{-1}$ values obtained from EUV photoevaporation alone and with the fact that in the EUV case the mass loss rate is more concentrated at a scale of a few AU.

  When combined with viscous evolution, X-EUV photoevaporation produces a qualitatively similar
pattern of inner hole creation followed by inside out clearing as is found in EUV models. The
initial size of the inner hole (about an AU) and the rapid clearing following gap opening is
similar in X-EUV and EUV models. The important difference is that this occurs earlier, and at much
higher accretion rate, in the X-EUV case. This implies that inner holes can be associated with more 
massive outer discs and with higher accretion rates on to the star than was possible in the EUV case.
Our results thus weaken the objections made previously to EUV photoevaporation with regard to
inner hole creation.  For example, the X-EUV model can open a gap even in
the case of massive outer discs such as that 
in TW Hydra. 

  Nevertheless there are generic predictions of photoevaporation models
(whether EUV or X-EUV) that appear not to be compatible with features of
some of the best studied inner hole sources. Namely: i) photoevaporation models
predict that gas and  dust in the inner hole, as well as diagnostics
of accretion onto the star, should be observable only during inner disc
clearing and that during this phase the ratio of inner disc gas mass to
accretion rate should be of order the lifetime of this phase. In
TW Hydra and GM Auriga, however, the observed ratio is very low and would
imply an unacceptably short lifetime of this phase. This low value
probably implies that, in contrast to what is predicted in photoevaporation
models, the gas and dust in the inner hole is continuously replenished
from the outer disc. ii) Photoevaporation models also predict that the
phase of inner hole draining is associated with small holes (i.e. of
$< 10$ AU scale). There is no mechanism to produce large holes that
contain gas and undergo vigorous accretion onto the star (as
in 
GM Auriga).

 On the other hand,
both EUV and X-EUV models can explain the detection  of outer discs in around
$10 \%$ of  stars
which show no evidence for   inner disc emission/accretion diagnostics
\citep{andrews05}. Future measurements
with Herschel or ALMA are likely to uncover  a class of objects with large
inner holes and we would expect this population to be
dominated by photoevaporative holes.

 Finally, we stress that the X-EUV case investigated here 
corresponds to the case of a rather large X-ray luminosity ($2\times 10^{30}$erg~s$^{-1}$) and further work
is required in order to explore the effect on disc evolution of the wide range of X-ray luminosities measured
in T Tauri stars. A clear qualitative prediction is that  disc lifetimes
are expected to be shorter in
X-ray luminous sources and that these sources should be be over-represented among disc-less
(Weak Line) T Tauri stars. This prediction is  in line with observations:
though the larger X-ray luminosity in Weak Line T Tauri stars is
conventionally ascribed to suppression of X-rays in accreting systems,
we here follow \citet{drake09} in suggesting that another factor is
that X-rays may instead suppress accretion.

\section*{ACKNOWLEDGMENTS}
We thank Paola D'Alessio for providing us with the electronic data for the gas density distribution in the disc. We would also like to thank the referee for their comments and suggested imporvements. JEO acknowledges support of a STFC PhD studentship. BE acknowledges funding from a STFC advanced fellowship. RDA acknowledges support from the Netherlands Organisation for Scientific Research (NWO), through VIDI grants 639.042.404 and 639.042.607. This work was partly performed using the Darwin Supercomputer of the University of Cambridge High Performance
Computing Service (http://www.hpc.cam.ac.uk/), provided by Dell Inc. using Strategic Research Infrastructure Funding from the Higher Education Funding Council for England.

\label{lastpage}

\end{document}